\documentclass{article}
\usepackage{fixltx2e}
\usepackage[pdftex]{hyperref}
\usepackage[nofiglist]{endfloat}

\newcommand{\lyxaddress}[1]{
\par {\raggedright #1
\vspace{1.4em} 
\noindent\par}
}

\advance\voffset -.5in 

\usepackage[pdftex]{graphicx}

\usepackage{amssymb,amsfonts,amsmath}

\newcommand{\tRNA}[2]{tRNA\rlap{\textsuperscript{#1}}\textsubscript{#2}}
\newcommand{\tRNAaa}[1]{tRNA\textsuperscript{#1}}
\newcommand{\tDNA}[2]{tDNA\rlap{\textsuperscript{#1}}\textsubscript{#2}}

\newcommand{\figcapone}{{\bf A universal schema for tRNA-protein interaction networks.}} 

\newcommand{\figcaptwo}{{\bf Function logos~\cite{Freyhult:2006dr} for
    two groups of alphaproteobacteria and overview of tRNA-CIF-based
    binary phyloclassification.}}

\newcommand{\figcapthree}{{\bf Leave-One-Out Cross-Validation
  (LOO-CV) scores of alphaproteobacterial genomes under two different
  binary phyloclassifiers. } A. tRNA-CIF-based phyloclassifier
  B. Total tRNA sequence-based phyloclassifer.}

\newcommand{\figcapfour}{{\bf Breakout of class contributions to scores under
  the tRNA CIF-based binary phyloclassifier.}}

\newcommand{\figcapfive}{{\bf Base compositions of alphaproteobacterial
  tRNAs showing convergence between Rickettsiales and
  SAR11. } A. Stacked bar graphs of tRNA base composition by
  clade. B. UPGMA clustering of clades based on Euclidean distances of
  tRNA base compositions under the centered log ratio
  transformation~\cite{aitchison1986statistical}.}

\newcommand{\figcapsix}{{\bf FastUniFrac-based phylogenetic tree of
  alphaproteobacteria using tRNA data computed according to the
  methods of~\cite{Widmann:2010ea}.}}

\newcommand{\figcapseven}{{\bf Seven-way tRNA-CIF-based
    phyloclassification of alphaproteobacterial genomes by the default
    multilayer perceptron in WEKA.} Bootstrap support values under
  resampling of tRNA sites against (left) all tRNA CIFs and (right)
  CIFs with Gorodkin heights $\ge 0.5$ bits and model retraining (100
  replicates). All support values correspond to most probable clade as
  shown except for {\it Stapphia} and {\it Labrenza} for which they
  correspond to Rhizobiales.  }

\begin{document}


\title {tRNA signatures reveal polyphyletic origins of streamlined SAR11 genomes among the alphaproteobacteria}

\author{Katherine C.H. Amrine$^{\text{1}}$, Wesley D. Swingley$^{\text{1,2}}$, David H. Ardell$^{\text{1,*}}$}

\maketitle

\lyxaddress{1. Program in Quantitative and Systems Biology, 5200 North Lake Road, University of California, Merced, Merced, CA 95343}
\lyxaddress{2. Current Address: Department of Biological Sciences, Northern Illinois University, DeKalb, IL 60115}
\lyxaddress{{*} Corresponding Author: David H. Ardell, (209) 228-2953, dardell@ucmerced.edu.}

\section*{Abstract}
Phylogenomic analyses are subject to bias from convergence in
macromolecular compositions and noise from horizontal gene transfer
(HGT). Accordingly, compositional convergence leads to contradictory
results on the phylogeny of taxa such as the ecologically dominant
SAR11 group of Alphaproteobacteria, which have extremely streamlined,
A+T-biased genomes. While careful modeling can reduce bias artifacts
caused by convergence, the most consistent and robust phylogenetic
signal in genomes may lie distributed among encoded functional
features that govern macromolecular interactions.  Here we develop a
novel phyloclassification method based on signatures derived from
bioinformatically defined tRNA Class-Informative Features (CIFs). tRNA
CIFs are enriched for features that underlie tRNA-protein
interactions. Using a simple tRNA-CIF-based phyloclassifier, we
obtained results consistent with bias-corrected whole proteome
phylogenomic studies, rejecting monophyly of SAR11 and affiliating
most strains with Rhizobiales with strong statistical support. Yet, as
expected by their elevated genomic A+T contents, SAR11 and
Rickettsiales tRNA genes are also similarly and distinctly A+T-rich
within Alphaproteobacteria. Using conventional supermatrix methods on
total tRNA sequence data, we could recover the artifactual result of a
monophyletic SAR11 grouping with Rickettsiales. Thus tRNA CIF-based
phyloclassification is more robust to base content convergence than
supermatrix phylogenomics with whole tRNA sequences. Also, given the
notoriously promiscuous HGT rates of aminoacyl-tRNA synthetase genes,
tRNA CIF-based phyloclassification may be at least partly robust to
HGT of network components. We describe how unique features of the
tRNA-protein interaction network facilitate mining of traits governing
macromolecular interactions from genomic data, and discuss why
interaction-governing traits may be especially useful to solve
difficult problems in microbial classification and phylogeny.


\section*{Author Summary}
In this study, we describe a new way to classify living things using
information from whole genomes. First, for a group of related
organisms, we bioinformatically predict features by which specific
classes of tRNAs are recognized by certain proteins or
complexes. Second, we train an artificial neural network to recognize
which code a new, unknown genome belongs to. We apply our method to
SAR11, one of the most abundant bacteria in the world's oceans. We
find that different strains of SAR11 are more distantly related, both
to each other and to mitochondria, than previously thought. However,
with more traditional treatments of whole tRNA sequence data, we
obtain different results, best explained as artifacts of base content
convergence. Our tRNA features are therefore more robust to genomic
base content convergence than the tRNAs in which they are embedded;
this is additional evidence of their functional importance. The tRNA
features we study form a clade-specific and slowly diverging ``feature
network'' that underlies a universally conserved macromolecular
interaction network. We discuss on theoretical grounds why traits
governing macromolecular interactions may be especially well-suited to
resolve deep relationships in the Tree of Life.

\section*{Introduction}

What parts of genomes are most robust to compositional convergence?
What information is most faithfully inherited vertically? The key
assumptions of compositional stationarity and consistency in gene
histories underpin most current approaches in phylogenomics and are
frequently violated (reviewed in {\it e.g.}\cite{Gribaldo:2002}). HGT
is so widespread that the very existence of a ``Tree of Life'' has
been questioned~\cite{Gogarten:2002us,Bapteste:2009ci}. Better
understanding of ancient phylogenetic relationships requires discovery
of new universal, slowly-evolving phylogenetic markers that are robust
to compositional convergence and HGT.

The controversial phylogeny of {\it Ca.}Pelagibacter ubique (SAR11) is
a case in point. SAR11 make up between a fifth and a third of the
bacterial biomass in marine and freshwater
ecosystems~\cite{Morris:2002bn}. Adaptations to extreme environmental
nutrient limitation may explain why SAR11 have very small cell and
genome sizes and small fractions of intergenic
DNA~\cite{Giovannoni:2005}. While some recent phylogenomic studies
define a clade among SAR11, the largely endoparasitic Rickettsiales,
and the alphaproteobacterial ancestor of
mitochondria~\cite{Williams:2007p5327,Georgiades:2011gx,Thrash:2011kl},
others argue that this placement of SAR11 is an artifact of
independent convergence towards increased genomic A+T content, and
that SAR11 belongs closer to other free-living Alphaproteobacteria
such as the Rhizobiales and
Rhodobacteraceae~\cite{Brindefalk:2011ei,RodriguezEzpeleta:2012fw,Viklund:2012jr}.
Monophyly of SAR11 was also recently
rejected~\cite{RodriguezEzpeleta:2012fw}.

Nonstationary macromolecular compositions are a known source of bias
in phylogenomics~\cite{Foster01062004,Losos14122012}. Widespread variation in
macromolecular compositions may be associated with loss of DNA repair
pathways in reduced genomes~\cite{Dale:2003hc,Viklund:2012jr},
unveiling an inherent A+T-bias of mutation in
bacteria~\cite{Hershberg:2010ig} and elevating genomic A+T
content~\cite{Moran:2002p5737,Lind:2008cs}. A process such as this has
likely altered protein and RNA compositions genome-wide in SAR11, and
if such effects are accounted for, the placement of SAR11 with
Rickettsiales drops away as an apparent
artifact~\cite{RodriguezEzpeleta:2012fw,Viklund:2012jr}. Consistent
with this interpretation, SAR11 strain HTTC1062 shares a surprising
and unique codivergence of \tRNAaa{His} and histidyl-tRNA synthetase
(HisRS) with a clade of free-living
Alphaproteobacteria~\cite{Wang:2007p1040,Ardell:2010du} that likely
arose only once in bacteria~\cite{Ardell:2006ko}. This synapomorphy
contradicts the placement of SAR11 with Rickettsiales.

\begin{figure*}
\centering
\includegraphics[width=3.2in]{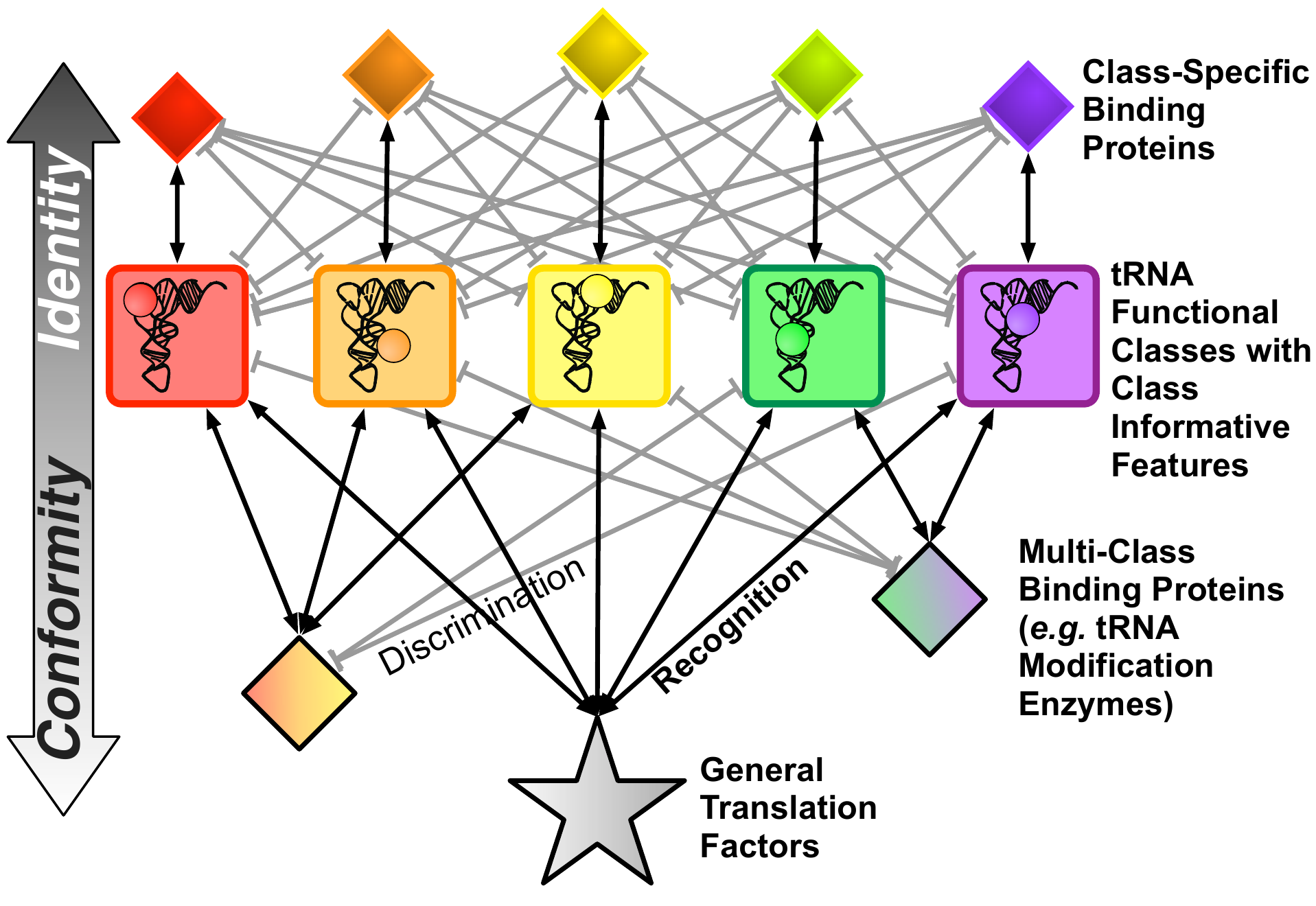}
\caption{\figcapone}
\label{fig:network}
\end{figure*}

This work was motivated to determine whether the entire system of
tRNA-protein interactions could be exploited to address phylogeny of
bacteria, particularly SAR11. The highly conserved tRNA-protein
interaction network (Fig.~\ref{fig:network}) has special advantages
for comparative systems biological study from genomic data. First, the
components and interactions of this network are highly
conserved. Second, bioinformatic mining of interaction-determining
traits from genomic tRNA data is favorable because tRNA structures are
highly conserved not just across extant taxa but also across different
functional classes of tRNAs
(``conformity''~\cite{WOLFSON01012001}). Yet each functional class of
tRNA must maintain a hierarchy of increasingly specific interactions
with various proteins and other factors
(``identity''~\cite{Giege:2008}). The conflicting requirements of
conformity and identity allow structural comparison and contrast to
predict class-informative traits of tRNAs from sequence data by
relatively simple bioinformatic methods~\cite{Ardell:2010du}. The
features that govern tRNA-protein interactions diverge across the
three domains of life (reviewed in~\cite{Giege:1998um}) and also
within the domain of bacteria~\cite{Ardell:2006ko}.

In prior work, we developed ``function logos'' to predict, at the
level of individual nucleotides before post-transcriptional
modification, what genetically templated information in tRNA gene
sequences is associated to specific functional identity
classes~\cite{Freyhult:2006dr}. We now call these function-logo-based
predictions Class-Informative Features (CIFs). A tRNA CIF answers a
question like: ``if a tRNA gene from a group of related genomes
carries a specific nucleotide at a specific structural position, how
much informaiton do we gain about that tRNAs specific function?''
Such information estimates are corrected for biased sampling of
functional classes and sample size effects~\cite{Freyhult:2006dr}, and
their statistical significance may be
calculated~\cite{Ardell:2010du}. Although an individual bacterial
genome does not present enough data to generate a function logo,
related genome data may be lumped, weakly assuming homogeneity of tRNA
identity rules (although heterogeneity generally reduces
signal). Function logos recover known tRNA identity elements ({\it
  i.e.} features that govern the specificity of interactions between
tRNAs and proteins)~\cite{Giege:1998um}, and more generally, predict
features governing interactions with class-specific network partners
such as amidotransferases~\cite{Bailly:2006cs}. A recent molecular
dynamics study on a \tRNAaa{Glu}-GluRS (Glutaminal tRNA-synthetase)
complex identified tRNA functional sites involved in intra- and
inter-molecular allosteric signalling within GluRS that couples
substrate recognition to reaction catalysis~\cite{Sethi:2009}. The
predicted sites are correlated with those from proteobacterial
function logos~\cite{Freyhult:2007jj}.

In this work, we show that tRNA CIFs have diverged among
Alphaproteobacteria in a phylogenetically informative manner. Second,
as phylogenetic markers, tRNA CIFs are more robust to compositional
convergence than the tRNA bodies in which they are embedded. Using our
tRNA-CIF-based phyloclassification approach, we confirm that SAR11 are
polyphyletic with the majority of strains clustering with the
free-living Alphaproteobacteria. Our results have implications for how
to best mine genomic data for phylogenetic signals.

\section*{Results}

We reannotated Alphaproteobacterial tDNA data from tRNAdb-CE
2011~\cite{Abe:2011js} and other prepublication genomic data, and
split them into two groups according to whether or not their source
genome contained the uniquely derived synapomorphic traits previously
described~\cite{Ardell:2006ko}: a gene for \tRNAaa{His} containing A73
(using ``Sprinzl coordinates'',~\cite{Sprinzl:1998vz}) and lacking
templated $-1G$. We could thereby partition the data into an RRCH
clade (Rhodobacteraceae, Rhizobiales, Caulobacterales,
Hyphomonadaceae), which present the uniquely derived \tRNAaa{His}, and
the RSR grade (Rhodospirillales, Sphingomonadales, and Rickettsiales,
excluding SAR11), which present ``normal'' bacterial \tRNAaa{His} with
C73 and genomically templated $-1G$.  In all, data from 214
Alphaproteobacterial genomes represented 11644 predicted tRNA
sequences (8773 sequences unique within genomes and 3064 total unique
sequences). Our final dataset contained 147 genomes (8597 tRNAs) for
the RRCH clade, 59 genomes (2792 tRNAs) for the RSR grade, and 8
genomes (255 tRNAs) of SAR11 strains.

\begin{figure*}
\centering
\includegraphics[width=5in]{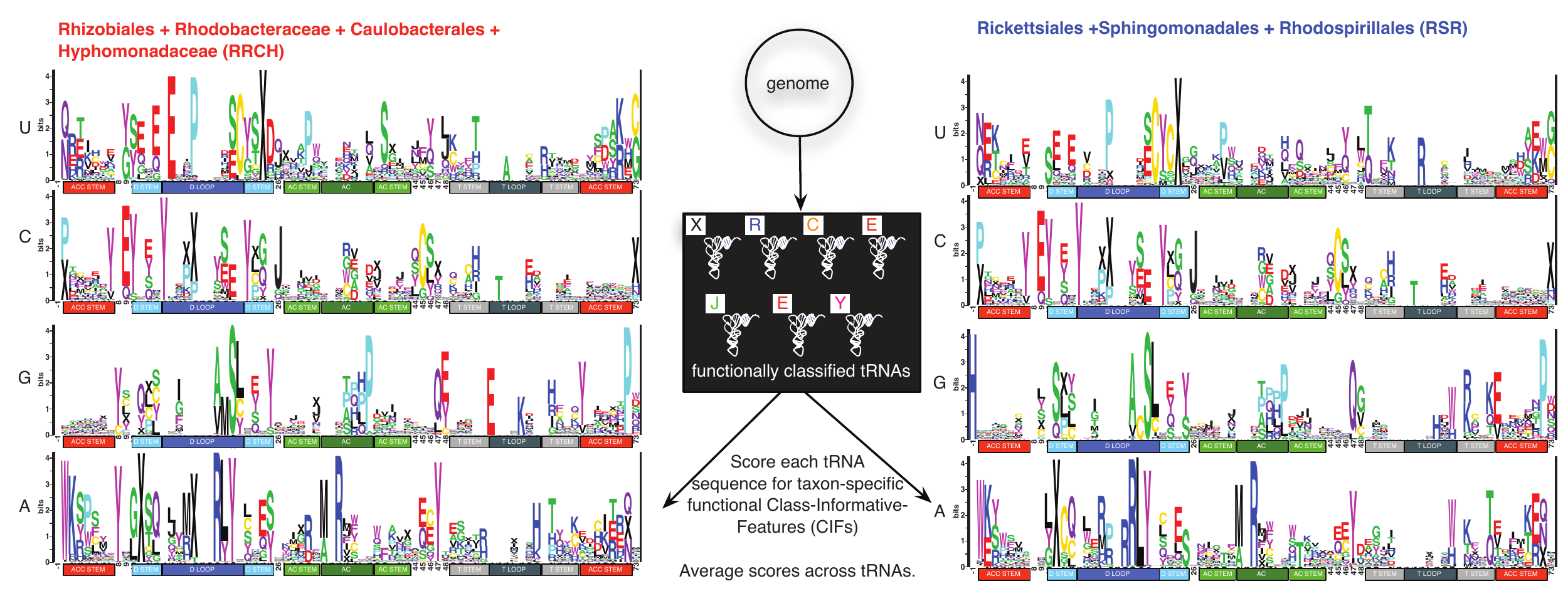}
\caption{\figcaptwo}
\label{fig:over}
\end{figure*}

The unique traits of the RRCH \tRNAaa{His} are perfectly associated to
substitutions of key residues in the motif IIb tRNA-binding loops of
HisRS involved in tRNA recognition~\cite{Ardell:2006ko}. Seven of
eight SAR11 strains exhibited the unique \tRNAaa{His}/HisRS
codivergence traits in common with RRCH genomes. In contrast, strain
HIMB59 presented ancestral bacterial characters in both \tRNAaa{His}
and HisRS (Fig. S1). These results immediately suggest that
HIMB59 is not monophyletic with the other SAR11 strains, consistent
with~\cite{RodriguezEzpeleta:2012fw}.

\begin{figure*}
\centering
\includegraphics[width=5in]{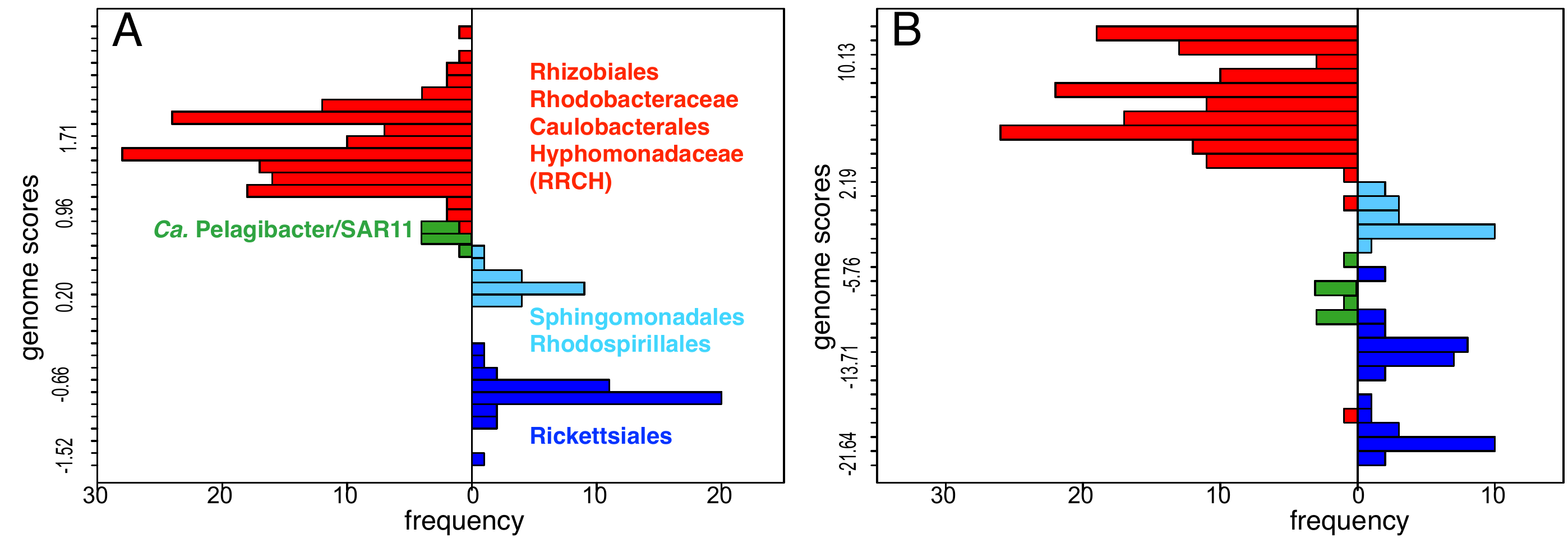} 
\caption{\figcapthree}
\label{fig:binclass}
\end{figure*}

We computed function logos~\cite{Freyhult:2006dr} of the RRCH clade
and RSR grade to form the basis of a tRNA-CIF-based binary
phyloclassifier as shown schematically in Fig.~\ref{fig:over}. To
reduce bias, we used a Leave-One-Out Cross-Validation (LOOCV)
approach. For comparison, we also performed LOOCV phyloclassification
using sequence profiles of entire tRNAs, with typical results shown in
Fig.~\ref{fig:binclass}B. Although the tRNA-CIF-based phyloclassifier
(Fig.~\ref{fig:binclass}A) was biased positively by the much larger
RRCH sample size, it achieved better phylogenetic separation of
genomes than the total-tRNA-sequence-based phyloclassifier
(Fig.~\ref{fig:binclass}B). The Sphingomonadales and Rhodospirillales
separated in scores from the Rickettsiales in both classifiers. Most
importantly, the tRNA-CIF-based phyloclassifier placed all eight SAR11
genomes closer to the RRCH clade and far away from the Rickettsiales
with HIMB59 overlapping the Rhodospirillales, while the
total-tRNA-sequence-based phyloclassifier placed all eight SAR11
genomes closer to the Rickettsiales. Fig.~S2 shows the effects of
different treatments of missing data in the total-tRNA-sequence-based
classifier. Method ``zero,'' shown in Fig.~\ref{fig:binclass}B, is
most analogous to the method used to generate
Fig.~\ref{fig:binclass}A. Method ``skip'' (Fig. S2B) shows that
SAR11 tRNAs share sequence characters in common with the RSR grade
that are not seen in the RRCH clade. Methods ``small'' and ``pseudo''
(Figs. S2C and S2D) show that SAR11 have sequence traits not
observed in either RSR or RRCH.

\begin{figure*}
\centering
\includegraphics[width=5in]{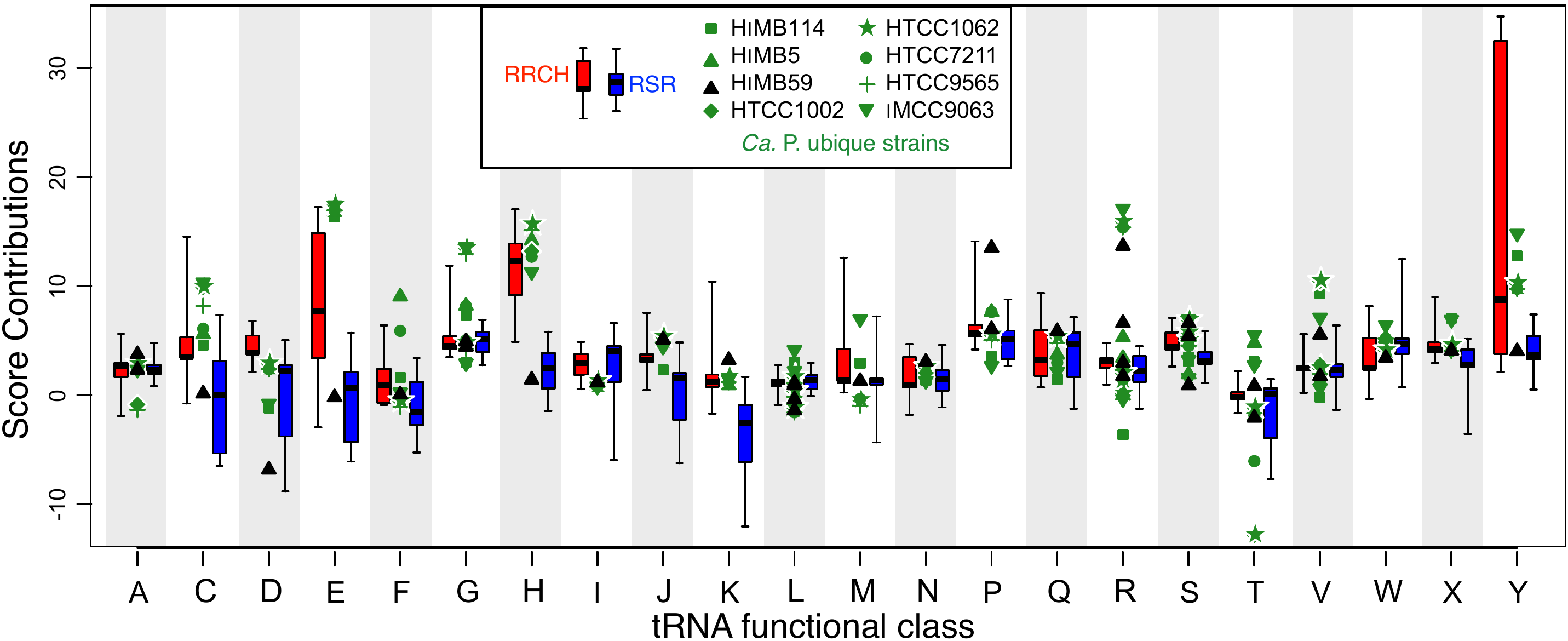}
\caption{\figcapfour}
\label{fig:breakout}
\end{figure*}

Many other tRNA classes besides \tRNAaa{His} contribute to the
differentiated classification of RRCH and RSR genomes by the CIF-based
binary classifier (Fig.~\ref{fig:breakout}). Other tRNA classes are
also differentiated between these two groups, including \tRNAaa{Cys},
\tRNAaa{Asp}, \tRNAaa{Glu}, \tRNA{Ile}{LAU} (symbolized ``J''),
\tRNAaa{Lys}, \tRNAaa{Tyr}. These results extend the observations of
\cite{Wang:2007p1040} who discovered unusual base-pair features of
\tRNAaa{Glu} in the RRCH clade. In classes for which the RRCH and RSR
groups are well-differentiated, HIMB59 uniquely groups with RSR while
other strains group with RRCH, while for other tRNA classes, all
putative SAR11 strains lie outside the RRCH and RSR
distributions. This implies that more diverse Alphaproteobacterial
genomic data are necessary to completely resolve the phylogenetic
affiliation of SAR11 strains, but strongly contradict a monophyletic
affiliation of SAR11 with Rickettsiales. 

\begin{figure*}
\centering
\includegraphics[width=5in]{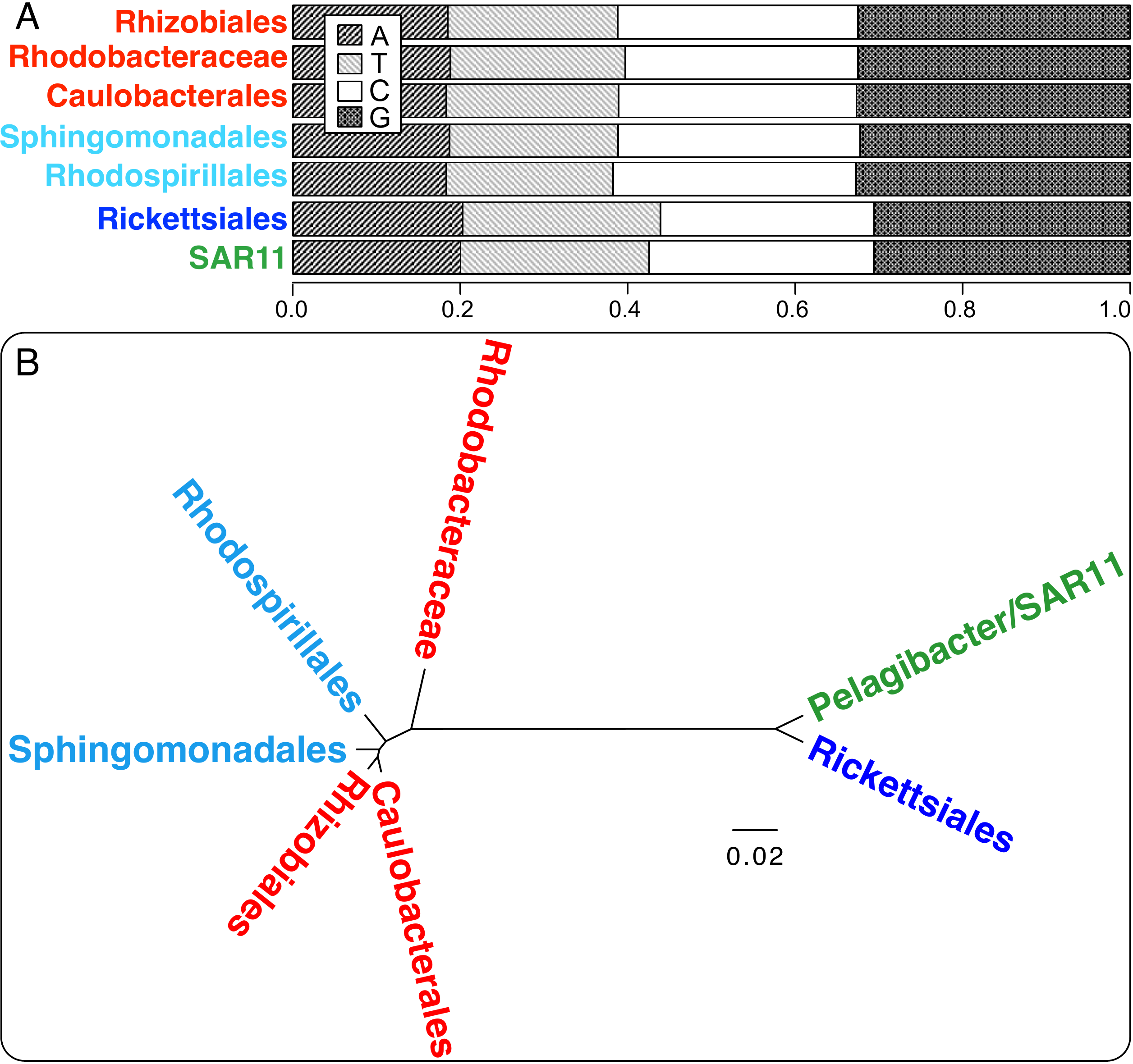}
\caption{\figcapfive}
\label{fig:comp}
\end{figure*}

The increases in genomic A+T contents in SAR11 and Rickettsiales have
also driven elevated A+T contents of their tRNA genes
(Fig.~\ref{fig:comp}A).  Rickettsiales and SAR11 tRNA genes are both
notably elevated in both A and T, and share an overall similarity in
composition distinct from other Alphaproteobacteria. Hierarchical
clustering of Alphaproteobacterial taxa based on tRNA gene base
contents closely group SAR11 and Rickettsiales together
(Fig.~\ref{fig:comp}B).

\begin{figure*}
\centering
\includegraphics[width=5in]{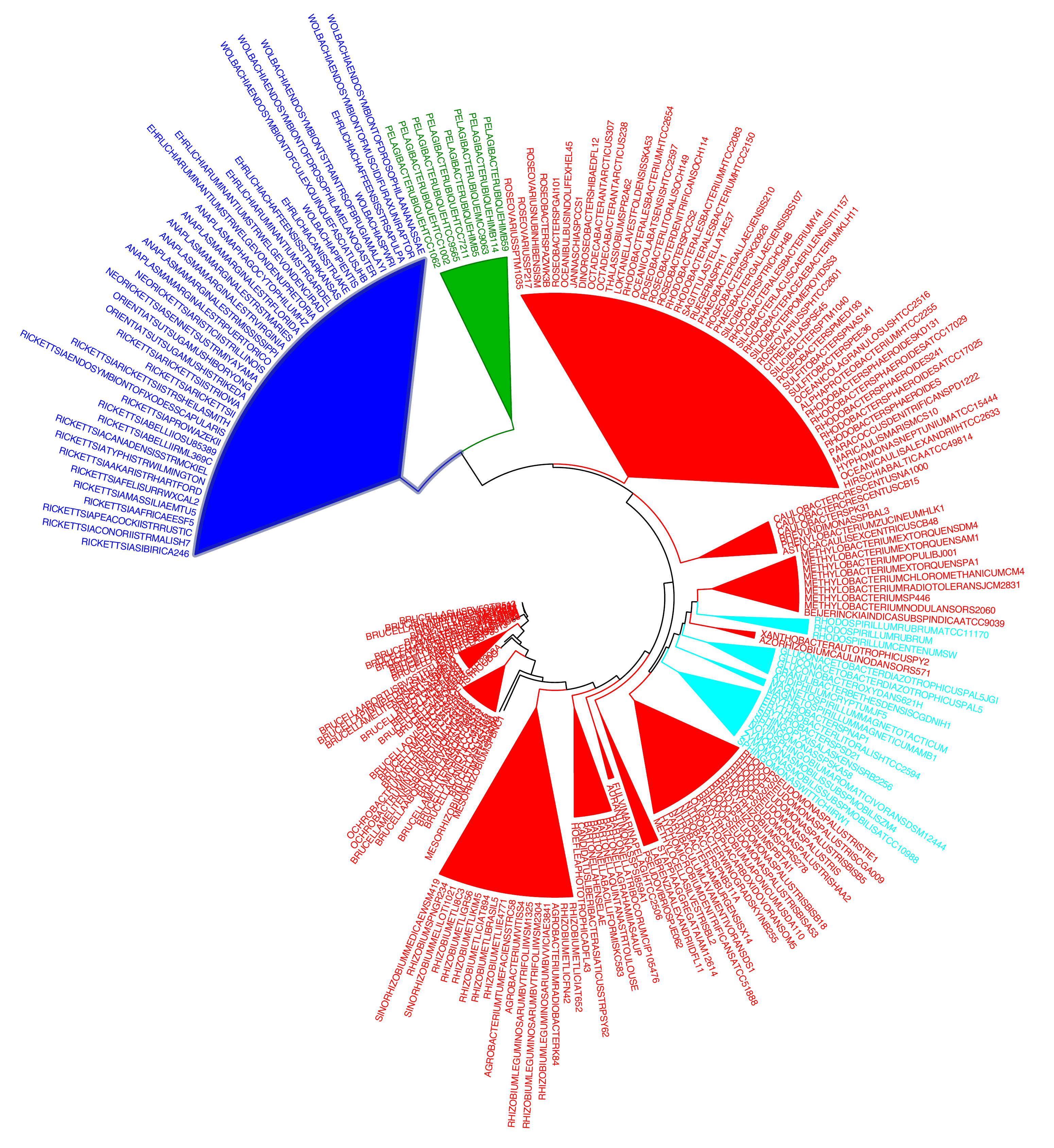}
\caption{\figcapsix}
\label{fig:unifrac}
\end{figure*}

Nonstationary tRNA base content --- convergence to greater A+T content
--- causes all eight SAR11 strains in our dataset to group with
Rickettsiales using phylogenomic approaches based on total tRNA
sequence evidence. In a ``supermatrix'' phylogenomic approach,
concatenating genes for 28 isoacceptor classes from 169 species (2156
total sites) and using the GTR+Gamma model in RAxML, we estimated a
Maximum Likelihood tree in which all eight putative SAR11 strains
branch together with Rickettsiales (Fig.~S3). For this analysis,
in 31\% of instances when isoacceptor genes were picked from a genome,
we randomly picked one gene from a set of isoacceptor
paralogs. However, our results did not depend on which paralog we
picked. Using a distance-based approach with FastTree, we computed a
consensus cladogram over 100 replicate alignments each representing
different randomized picks over paralogs. As the consensus cladogram
shows (Fig. S4) each replicate distance tree placed all
eight putative SAR11 strains together with Rickettsiales. The recently
introduced tRNA-specific FastUniFrac-based method for microbial
classification~\cite{Widmann:2010ea} also places all SAR11 strains
together with Rickettsiales (Fig.~\ref{fig:unifrac}).

\begin{figure*}
\centering
\includegraphics[width=6in]{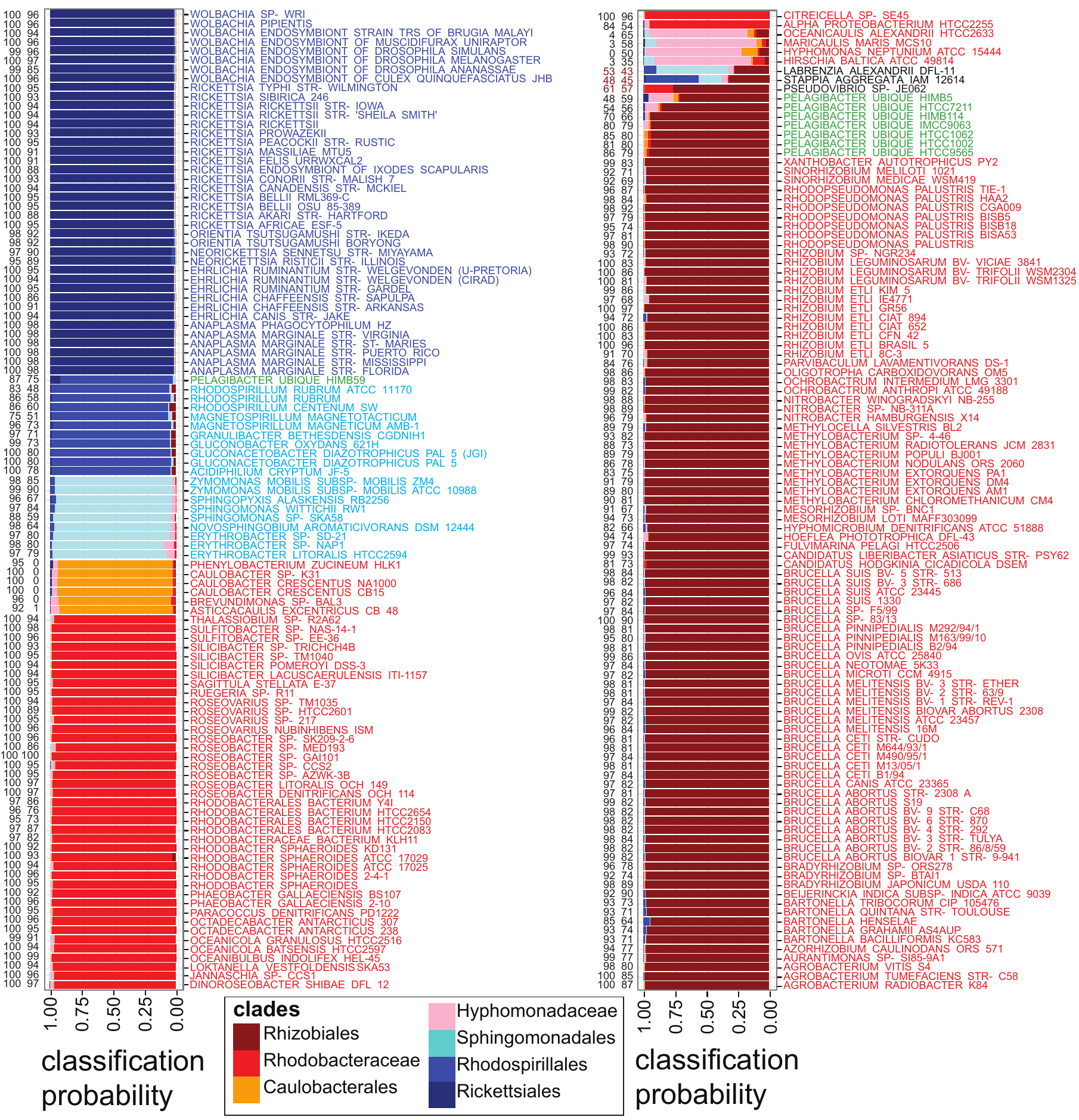}
\caption{\figcapsix}
\label{fig:multiway}
\end{figure*}

However, as shown in Fig.~\ref{fig:multiway}, a multiway classifier
based on tRNA CIFs bins all SAR11 strains with the Rhizobiales except
for HIMB59, which bins with the Rhodospirillales, consistent with the
results of~\cite{RodriguezEzpeleta:2012fw}. These results use a
multilayer perceptron (MLP) classifier implemented in
WEKA~\cite{Hall:2009ud} and only seven taxon-specific CIF-based
summary scores. The MLP is the simplest non-linear classifier able to
handle the interdependent signals in the CIF-based scores for
tree-like data~\cite{duda2012pattern}. In a Leave-One-Out
cross-classification, all other genomes scored consistently with NCBI
Taxonomy except three placed in Rhodobacteraceae based on 16S
ribosomal RNA evidence: {\it Stappia aggregata}, {\it Labrenzia
  alexandrii} and the denitrifying {\it Pseudovibrio sp.} JE062. None
of these genomes scored strongly against Rhodobacteraceae except {\it
  Pseudovibrio}, which scored four times greater against the
Rhizobiales. 

To assess robustness of our results we performed two controls: we
bootstrapped sites of tRNA data in each genome to be classified, and
we filtered away small CIFs with Gorodkin heights $< 0.5$ bits from
our models, retrained the classifier and bootstrapped sites
again. Generally bootstrap support values correspond to original
classification probabilities. All SAR11 strains have support values
$>80\%$ as Rhizobiales, majority bootstrap values as Rhizobiales
(HIMB114 at $70\%$ with Rickettsiales at $15\%$ and HTCC7211 at $54\%$
with Rickettsiales at $13\%$), or plurality bootstrap value as
Rickettsiales (HIM5 at $48\%$ with Rickettsiales at $18\%$) except
HIM59 which had a bootstrap support value of $87\%$ to be in the
Rhodospirillales. Full bootstrap statistics with these model are
provided in Table S1.

\section*{Discussion}

Our results provide strong, albeit unconventional, evidence that most
SAR11 strains are affiliated with Rhizobiales, while strain HIMB59 is
affiliated with Rhodospirillales.  These results are entirely
consistent with comprehensive phylogenomic studies that control for
nonstationary macromolecular compositions in
Alphaproteobacteria~\cite{Brindefalk:2011ei,RodriguezEzpeleta:2012fw,Viklund:2012jr}
or a site-rate-filtered analysis~\cite{Gupta:2007}. Our CIF-based
method works even though SAR11 and Rickettsiales tRNAs have converged
in base content, so that total tRNA sequence-based phylogenomics gives
opposite results.  tRNA CIFs must be at least partly robust to
compositional convergence of the tRNA bodies in which they are
embedded.

It is well known that aminoacyl-tRNA synthetases (aaRS) are highly
prone to
HGT~\cite{Doolittle:1998bn,Brown:1999ur,Wolf:1999vo,Woese:2000uy,Andam:2011gh}
including in
Alphaproteobacteria~\cite{Ardell:2006ko,Dohm:2006ts,Brindefalk:2006jda}. We
hypothesize that our tRNA-CIF-based phyloclassifiers are also robust
to HGT of components of the tRNA-protein interaction network,
consistent with \cite{Shiba:1997te}, who argued that a horizontally
transferred aaRS is more likely to functionally ameliorate to a
tRNA-protein network into which it has been transferred rather than
remodel that network to accomodate itself. HGT of aaRSs may also
perturb a network so as to cause a distinct pattern of divergence
(\cite{Ardell:2006ko} and this work). Wang {\it et
  al.}~\cite{Wang:2007p1040} discuss the possibility that RRCH
\tRNAaa{His} and HisRS were co-transferred into an ancestral SAR11
genome. However, this fails to explain the correlations of many other
tRNA traits of SAR11 genomes with the RRCH clade reported
here. Further study is needed to address the robustness of our method
to component HGT.

A more distant relationship between most SAR11 strains and
Rickettsiales actually strengthens the genome streamlining
hypothesis~\cite{Giovannoni:2005}. If SAR11 were a true branch within
Rickettsiales, it becomes more difficult to claim that genome
reduction in SAR11 occurred by a selection-driven evolutionary process
distinct from the drift-dominated erosion of genomes in the
Rickettsiales~\cite{Andersson:1998p7319,Moran:2002p5737,Itoh:2002dr}. By
the same token, polyphyly of nominal SAR11 strains implies that the
extensive similarity in genome structure and other traits between
HIMB59 and SAR11 reported by~\cite{Grote:2012ge} may have originated
independently. Perhaps convergence in some traits is consistent with
streamlining, which could also explain trait-sharing between SAR11 and
{\it Prochlorococcus}, marine cyanobacteria also argued to have
undergone streamlining~\cite{Dufresne:2005gn}. Clear signs of
data-limitation in our study should be taken to mean that better
taxonomic sampling will improve our results and could ultimately
resolve more than two origins of SAR11-type genomes among
Alphaproteobacteria.

We extracted accurate and robust phylogenetic signals from tRNA gene
sequences by first integrating within genomes to identify features
likely to govern functional interactions with other
macromolecules. Unlike small molecule interactions, macromolecular
interactions are mediated by genetically determined structural and
dynamic complementarities. These are intrinsically relative; a large
{\it neutral network}~\cite{Schuster:1994gf} of
interaction-determining features should be compatible with the same
interaction network. Coevolutionary divergence --- turnover---of
features that mediate macromolecular interactions, while conserving
network architecture, has been described in the transcriptional
networks of yeast~\cite{Kuo:2010fk,Baker:2011fz} and
worms~\cite{Barriere:2012dp} and in post-translational modifications
underlying protein-protein interactions~\cite{Krogan:2012cs}.  This
work demonstrates that divergence of interaction-governing features is
phylogenetically informative.

It remains open how such features diverge, with possibilities
including compensatory nearly neutral mutations~\cite{Hartl:1996dy},
fluctuating selection~\cite{He:2011jg}, adaptive
reversals~\cite{Bullaughey:2012dg}, and functionalization of
pre-existent variation~\cite{Haag:2005ty}. Major changes to
interaction interfaces may be sufficient to induce genetic isolation
between related lineages, as discussed for the 16S rRNA- and 23S
rRNA-based standard model of the ``Tree of Life,'' in which many
important and deep branches associate with large, rare macromolecular
changes (``signatures'') in ribosome structure and
function~\cite{Winker:1991jd,Roberts:2008di,Chen:2010et}.

Interaction-mediating features of macromolecules may be systems
biology's answer to the phylogeny problem. Perhaps no other traits of
genomes are vertically inherited more consistently than those that
mediate functional interactions with other macromolecules in the same
lineage. In fact, the structural and dynamic basis of interaction
among macromolecular components --- essential to their collaborative
function in a system --- may define a lineage better than any of those
components can themselves, either alone or in ensemble.

\section*{Materials and Methods}

Supplementary data packages are provided to reproduce all figures from
raw data and enable third-party classification of alphaproteobacterial
genomes.

\subsection*{Data}
The 2011 release of the tRNAdb-CE
database~\cite{Abe:2011js} was downloaded on August 24,
2011.  From this master database, we selected Alphaproteobacteria data
as specified by NCBI Taxonomy data (downloaded September 24, 2010,
\cite{Sayers:2010p4006}). Also using NCBI Taxonomy, we further
tripartitioned Alphaproteobacterial tRNAdb-CE data into those from the
RRCH clade, the RSR grade (excluding SAR11), and three SAR11
genomes, as documented in Supplementary data for figure 2. Five
additional SAR11 genomes (for strains HIMB59, HIMB5, HIMB114, IMCC9063
and HTCC9565) were obtained from J. Cameron Thrash courtesy of the lab
of S. Giovannoni. We custom annotated tRNA genes in these genomes as
the union of predictions from tRNAscan-SE version 1.3.1 (with {\tt -B}
option, \cite{LoweEddy97}) and Aragorn version
1.2.34~\cite{Laslett:2004ih}. We classified initiator tRNAs and
\tRNA{Ile}{CAU} using TFAM version 1.4~\cite{Taquist:2007jl} using a
model previously created to do this based on identifications
in~\cite{Silva:2006jc} provided as supplementary data.  We aligned
tRNAs with covea version 2.4.4~\cite{Eddy:1994ul} and the prokaryotic
tRNA covariance model~\cite{LoweEddy97}, removed sites with more than
97\% gaps with a bioperl-based utility~\cite{Stajich:2002}, and edited
the alignment manually in Seaview 4.1~\cite{gouy2010seaview} to remove
CCA tails and remove sequences with unusual secondary structures.  We
mapped sites to Sprinzl coordinates manually~\cite{Sprinzl:1998vz} and
verified by spot-checks against tRNAdb~\cite{juhling2009trnadb}.  We
added a gap in the -1 position for all sequences and -1G for
\tRNAaa{His} in the RSR group~\cite{Wang:2007p1040}.

\subsection*{tRNA CIF Estimation and Binary Classifiers}
Our tRNA-CIF-based binary phyloclassifier with Leave-One-Out
Cross-Validation (LOO CV) is computed directly from function logos,
estimated from tDNA alignments as described in~\cite{Freyhult:2006dr}.
Here, we define a {\it feature} $f \in F$ as a nucleotide $n \in N$ at
a position $l \in L$ in a structurally aligned tDNA, where $N =
\{A,C,G,T\}$ and $L$ is the set of all Sprinzl
coordinates~\cite{Sprinzl:1998vz}. The set $F$ of all possible
features is the Cartesian product $F = N \times L$.  A {\it functional
  class} or {\it class} of a tDNA is denoted $c \in \mathcal{C}$ where
$\mathcal{C} = \{A,C,D,E,F,G,H,I,J,K,L,M,N,P,Q,R,S,T,V,W,X,Y\}$ is the
universe of functions we here consider, symbolized by IUPAC one-letter
amino acid codes (for aminoacylation classes), $X$ for initiator
tRNAs, and $J$ for \tDNA{Ile}{LAU}. A {\it taxon set of genomes} or
just {\it taxon set} $S \in \mathcal{P}(G)$ is a set of genomes, where
$G$ is the set of all genomes, and $\mathcal{P}(G)$ is the power set
of $G$. In this work a genome $G$ is represented by the multiset of
tDNA sequences it contains, denoted $T_G$. The functional information of
features is computed with a map $h:(F \times C \times \mathcal{P}(G))
\longrightarrow \mathbb{R}_{\geq 0}$ from the Cartesian product of
features, classes and taxon sets to non-negative real numbers. For a
feature $f \in F$, class $c \in \mathcal{C}$ and taxon set $S \in
\mathcal{P}(G)$, $h(f,c,S)$ is the fraction of functional information
or ``Gorodkin height''~\cite{Gorodkin:1997tv}, measured in bits,
associated to that feature, class and taxon set. In this work, for a
given taxon set $S$, a function logo $H(S)$ is the tuple:

\begin{equation}
H(S) = \{(\alpha,\beta) \mid  \beta = h(\alpha, S), \forall \alpha \in (F \times \mathcal{C})\}.
\end{equation}

Furthermore the set $I(S) \subset (F \times \mathcal{C})$ of {\it tRNA
  Class-Informative Features} for taxon set $S$ is defined:

\begin{equation}
I(S) = \{\alpha \in (F \times \mathcal{C}) \mid h(\alpha, S) > 0\}.
\end{equation}

Briefly, a tRNA Class-Informative Feature is a tRNA structural feature
that is informative about the functional classes it associates with,
given the context of tRNA structural features that actually co-occur
among a taxon set of related cells, and corrected for biased sampling
of classes and finite sampling of
sequences~\cite{Freyhult:2006dr}. Let $A$ denote a set of
Alphaproteobacterial genomes partitioned into three disjoint subsets
$X$, $Y$ and $Z$ with $X \cup Y \cup Z = A$, representing genomes from
the RRCH clade, the RSR grade, and the eight nominal {\it Ca.}
Pelagibacter strains respectively. To execute Leave-One-Out
Cross-Validation of a tRNA CIF-based binary phyloclassifier for a
genome $G \in A$, we compute a score $S_C(G, S_1, S_2)$, averaging
contributions from the multiset $T_G$ of tDNAs in $G$ scored against
two function logos $H(S_1)$ and $H(S_2)$ computed respectively from
two disjoint taxon sets $S_1 \subset A$ and $S_2 \subset A$, with $G
\notin S_1 \cup S_2$. In this study, those sets are $X \setminus G$
and $Y \setminus G$, denoted $X_G$ and $Y_G$ respectively. Each tDNA
$t \in T_G$ presents a set of features $F_t \subset F$ and has a
functional class $c_t \in \mathcal{C}$ associated to it. The score
$S_C(G,X_G,Y_G)$ is then defined:

\begin{equation}
S_C(G,X_G, Y_G) \equiv \frac{1}{| T_G |}\sum_{t \in T_G}\sum_{f \in
  F_t}h(f,c_t,X_G) - h(f,c_t,Y_G).
\label{eqn:cif}
\end{equation}

As controls, we implemented four total-tDNA-sequence based binary
phyloclassifiers to score a genome $G$. All are slight variations in
which a tRNA $t \in T_G$ of class $c(t)$ contributes a score that is a
difference in log relative frequencies of the features it shares in
class-specific profile models generated from $X_G$ and $Y_G$. The
default ``zero'' scoring scheme method $S^Z_T(G,X_G, Y_G)$ shown in
Fig.~\ref{fig:binclass}B is defined as:

\begin{equation}
S^Z_T(G,X_G, Y_G) \equiv \frac{1}{| T_G |} \sum_{t \in T_G}\sum_{f \in
  F_t}\log_2 \frac{p^*(f | c_t,  X_G)}{p^*(f  | c_t, Y_G)}, 
\label{weaksol}
\end{equation}

\noindent where 

\begin{eqnarray}
p^*(f | c, S) \equiv 
\begin{cases} 
\#\{f , c, S\} / \#\{c,S\}  & \#\{f , c, S \} > 0 \\ 
1 & \#\{f , c, S \} = 0  
\end{cases},
\end{eqnarray}

\noindent $\#\{f, c, S \}$ is the observed frequency of feature $f$ in
tDNAs of class $c$ in set $S$, and $\#\{c, S \}$ is the frequency of
tDNAs of class $c$ in set $S$. 

Method ``skip'' corresponds to scoring scheme $S^K_T(G,X_G,
Y_G)$ defined as:

 \begin{equation}
S^K_T(G,X_G, Y_G) \equiv \frac{1}{| T_G |} \sum_{t \in T_G}\sum_{f \in F_t} s^k(f,c_t,  X_G, Y_G),
\end{equation}

\noindent where 

\begin{eqnarray}
s^k(f, c,  S, T) \equiv 
\begin{cases} 
\log_2 \frac{p(f | c, S)}{p(f | c, T)} & \#\{f , c, S \} > 0 \wedge \#\{f , c, T \} > 0 \\
 0 & \#\{f , c, S \} = 0 \vee \#\{f , c, T \} = 0  \end{cases},
\end{eqnarray}

\noindent and $p(f | c, R) \equiv \#\{f , c, R \}/ \#\{c,R\}$ for $R
\in \{S,T\}$ as before.



Methods ``pseudo'' and ``small'' correspond to scoring schemes $S^I_T(G,X_G,
Y_G)$:

 \begin{equation}
S^I_T(G,X_G, Y_G) \equiv 
 \frac{1}{| T_G |} \sum_{t \in T_G}\sum_{f \in F_t} \log_2
 \frac{p^I(f | c_t, X_G)}{p^I(f  | c_t, Y_G)},
\end{equation}

\noindent where 

\begin{eqnarray}
p^I(f | c, S) \equiv 
\begin{cases} o / t  &  \forall n \in N: \#\{(n,l) , c, S \} > 0  \\  
\frac{o + I}{t + 4I} & \exists n \in N: \#\{(n,l) , c, S \} = 0
\end{cases},
\end{eqnarray}

\noindent where $f = (n,l)$, $o \equiv \#\{f, c, S \}$, $t \equiv \#\{c,S\}$,
$I = 1$ for method ``pseudo,'' and, for method ``small,'' $I = 1 /
T_A$, where $T_A = \sum_{G \in A} T_G$.

\subsection*{Analysis of tRNA Base Composition}
We computed the base composition of tRNAs aggregated by clades using
bioperl-based~\cite{Stajich:2002} scripts, and transformed them by the
centered log ratio transformation~\cite{aitchison1986statistical} with
a custom script provided as supplementary data. We then computed
Euclidean distances on the transformed composition data, and then
performed hierarchical clustering by UPGMA on those distances as implemented in
the program NEIGHBOR from Phylip 3.6b~\cite{PHYLIP} and visualized in
FigTree v.1.4.

\subsection*{Supermatrix and FastUniFrac Analysis}
For supermatrix approaches, we created concatenated tRNA alignments
from 169 Alphaproteobacteria genomes (117 RRCH, 44 RSR, 8 PEL) that
all shared the same 28 isoacceptors with 77 sites per gene (2156 total
sites).  In cases where a species contained more than a single
isoacceptor, one was chosen at random.  Using a GTR+Gamma model, we
ran RAxML by means of The iPlant
Collaborative project RAxML server
(\url{http://www.iplantcollaborative.org}, \cite{Stamatakis01102008})
on January 23, 2013 with their installment of RAxML version
7.2.8-Alpha (executable raxmlHPC-SSE3, a sequential version of RAxML
optimized for parallelization).  We tested the robustness of our
result to random picking of isoacceptors by creating 100 replicate
concatenated alignments and running them through
FastTree~\cite{Price:2010vy}. For the FastUniFrac
analysis we used the FastUniFrac~\cite{Hamady:2010dr}
web-server at \url{http://bmf2.colorado.edu/fastunifrac/} to
accomodate our large dataset. We removed two genomes from our dataset
for containing fewer than 20 tRNAs, and following~\cite{Widmann:2010ea}
removed anticodon sites.  Following~\cite{Widmann:2010ea}
deliberately, we computed an approximate ML tree based on Jukes-Cantor
distances using FastTree~\cite{Price:2010vy}. We then
queried the FastUniFrac webserver with this tree, defining
environments as genomes. We then computed a UPGMA tree based on the
server's output FastUniFrac distance matrix in NEIGHBOR from Phylip
3.6b~\cite{PHYLIP}.

\subsection*{Multiway Classifier}
All tDNA data from the RSR and RRCH clades were partitioned into one
of seven monophyletic clades: orders Rickettsiales (N = 40 genomes),
Rhodospirillales (N = 10), Sphingomonadales (N = 9), Rhizobiales (N =
91), and Caulobacterales (N = 6), or families Rhodobacteraceae (N =
43) or Hyphomonadaceae (N = 4) as specified by NCBI taxonomy
(downloaded September 24, 2010,~\cite{Sayers:2010p4006}) and
documented in supplementary data for figure 7. We withheld data from
the eight nominal SAR11 strains, as well as from three genera {\it
  Stappia}, {\it Pseudovibrio}, and {\it Labrenzia}, based on
preliminary analysis of tDNA and CIF sequence variation. Following a
related strategy as with the binary classifier, we computed, for each
genome, seven tRNA-CIF-based scores, one for each of the seven
Alphaproteobacterial clades as represented by their function logos,
using the principle of Leave-One-Out Cross-Validation (LOO CV), that
is, excluding data from the genome to be scored. Function logos were
computed for each clade as described in~\cite{Freyhult:2006dr}.  For
each taxon set $X_G$ (with genome $G$ left out if it occurs), genome
$G$ obtains a score $S^M(G,X_G)$ defined by:

\begin{equation}
S_M(G,X_G) \equiv \frac{1}{| T_G |}\sum_{t \in T_G}  \sum_{f \in
  F_t}h(f,c_t,X_G).
\label{eqn:multi}
\end{equation}

Each genome $G$ is then represented by a vector of seven scores, one
for each taxon set modeled. These labeled vectors were then used to
train a multilayer perceptron classifier in WEKA 3.7.7~(downloaded
January 24, 2012,~\cite{Hall:2009ud}) by their defaults through the
command-line interface, which include a ten-fold cross-validation
procedure. We bootstrap resampled sites in genomic tRNA alignment data
(100 replicates) and also bootstrap resampled a reduced (and
retrained) model including only CIFs with a Gorodkin height~\cite{Freyhult:2006dr} $\ge
0.5$ bits.

\section*{Acknowledgments}
  We thank J. Cameron Thrash and Stephen Giovannoni for sharing data
  in advance of publication, Harish Bhat, Torgeir Hvidsten, Carolin
  Frank and Suzanne Sindi for helpful suggestions.

\bibliographystyle{plos2009}
\bibliography{MSB,KHA,manual}


\newpage{}

\section*{Figure Legends}

\paragraph*{Figure~1 }
\figcapone

\paragraph*{Figure~2 }
\figcaptwo

\paragraph*{Figure~3 }
\figcapthree

\paragraph*{Figure~4 }
\figcapfour

\paragraph*{Figure~5}
\figcapfive

\paragraph*{Figure~6}
\figcapsix

\paragraph*{Figure~7}
\figcapseven

\end{document}